\documentclass[%
 reprint,
 amsmath,amssymb,
 aps,
]{revtex4-2}
\usepackage[T1]{fontenc}
\usepackage{graphicx}
\usepackage{dcolumn}
\usepackage{bm}
\usepackage{hyperref}
\usepackage{xcolor}

\begin{document}

\preprint{APS/123-QED}



\title{Switching from a continuous to discontinuous phase transition under the quenched disorder.}

\author{Bartłomiej Nowak}
  \email{bartlomiej.nowak@pwr.edu.pl}
\author{Katarzyna Sznajd-Weron}%
  \email{katarzyna.weron@pwr.edu.pl}
\affiliation{%
Department of Theoretical Physics, Wrocław University of Science and Technology, 50-370 Wrocław, Poland
}%

\date{\today}

\begin{abstract}
Discontinuous phase transitions occurs to be particularly interesting from a social point of view  because of their relationship to social hysteresis and critical mass. In this paper, we show that the replacement of a time-varying  (annealed, situation-based) disorder by a static (quenched, personality-based) one can lead to a change from a continuous to a discontinuous phase transition. This is a result beyond the state of art, because so far numerous studies on various complex systems (physical, biological and social) have indicated that the quenched disorder can round or destroy the existence of a discontinuous phase transition. To show the possibility of the opposite behavior, we study a multistate $q$-voter model, with two types of disorder related to random competing interactions (conformity and anticonformity). We confirm, both analytically and through Monte Carlo simulations, that indeed discontinuous phase transitions can be induced by a static disorder.
\end{abstract}

\maketitle

\section{Introduction}
The study of complex systems, for which the 2021 Nobel Prize was awarded, is arguably the most interdisciplinary field of science, influencing many seemingly unrelated disciplines. For example, it may seem that physics and the social sciences have little in common, yet a huge number of papers have been published in recent decades using statistical physics methods to model various social systems \cite{Cas:For:Lor:09,Kwa:Dro:12,Per:etal:17,Jus:etal:22}. 

Someone might ask why physicists are concerned with social systems. Probably the first answer that comes to mind is that the methods and concepts of statistical physics can also be useful in the social sciences. But is feedback possible? Can problems and concepts from the social sciences trigger the development of physics itself? The very birth of statistical physics shows that this is what can happen \cite{Bal:02}. However today, as sometimes claimed, \textit{physicists are completing the circle by applying physical methods (often times those of statistical physics) to quantify social phenomena} \cite{Jus:etal:22}. This is undeniably true, but does this mean that nowadays research at the intersection of physics and social science no longer contributes anything new to physics?

In this work, we show that the result obtained within the model originally proposed to describe social opinion dynamics can go beyond the state of the art in physics. This result will address the effects of two types of approaches, so-called quenched and annealed, on phase transitions. Before we get to the point, let's clarify the terms \textit{quenched} and \textit{annealed} in the context of disorder in complex systems, because, in our experience, they are not widely known to the general audience. Within the quenched approach various randomness (heterogeneity) associated with interactions, topology, etc. are fixed in time. On the other hand, within the annealed approach all these randomness are changed at each time step \cite{Der:Sta:86}. 
In the context of social systems, inspired by the long-standing person-situation debate \cite{Luc:Don:09},  we related quenched to the personality-oriented, while annealed to the situation-oriented approach \cite{Szn:Szw:Wer:14,Jed:Szn:17}. Therefore, the analysis of differences resulting from the given approach (quenched vs annealed) is interesting in the context of social systems. But is this topic of interest in the context of physics itself? Looking at the literature it definitely does.

The role of the quenched disorder in shaping the type of the phase transitions (PTs) have been intensively studied from experimental and theoretical point of view, and applied to understand the behavior of various complex systems \cite{Voj:14}. In particular, it has been found that in low-dimensions quenched randomness results in rounding, smearing or completely destroying discontinuous  PTs \cite{Imr:Ma:75,Aiz:Weh:89,Bin:83,Voj:06,Mun:etal:10,Odo:Sim:21}. The early prediction of this effect was given heuristically by Imry and Ma \cite{Imr:Ma:75}, and later proven by Aizenman and Wehr \cite{Aiz:Weh:89}. More recently, it occurred to be true also in genuinely non-equilibrium systems \cite{Bor:Mar:Mun:13,Vil:Bon:Mun:14}. Another possibility is that discontinuous (mixed order) PT remains discontinuous and the heterogeneity adds a Griffiths phase subcritically \cite{Odo:Sim:21}. Moreover, for higher dimensional systems (three-dimensional or in the mean-field limit), it has been shown that a discontinuous phase transition can simply remain discontinuous in the presence of quenched disorder \cite{Fer:etal:08,Jed:Szn:17,Now:Sto:Szn:21}. However, to the best of our knowledge, the complementary effect, i.e., change from continuous to discontinuous PT under the quenched disorder has never been  observed, even in the mean-field limit.

In this paper we will show, both analytically and by Monte Carlo simulations, that such an effect is possible, at least on a complete graph (CG), which corresponds to the mean-field approach (MFA). We will show this within the framework of a model for which so far only the typical destruction (or softening) of discontinuous PT under quenched disorder has been observed, namely the $q$-voter model \cite{Cas:Mun:Pas:09}. For example, it was shown that the quenched disorder rounds discontinuous PT in the multistate $q$-voter model with independence \cite{Now:Sto:Szn:21} or completely kills these transitions in the two-state version of this model \cite{Jed:Szn:17}. Additionally, it was shown that in the two-state $q$-voter model with random competing interactions (conformity and anticonformity), both quenched and annealed disorder give exactly the same continuous PTs \cite{Jed:Szn:17}. The multistate version of such a model has not been studied yet and will be the subject of this paper.  

\section{Methods}
\begin{figure*}
  \includegraphics[width=\linewidth]{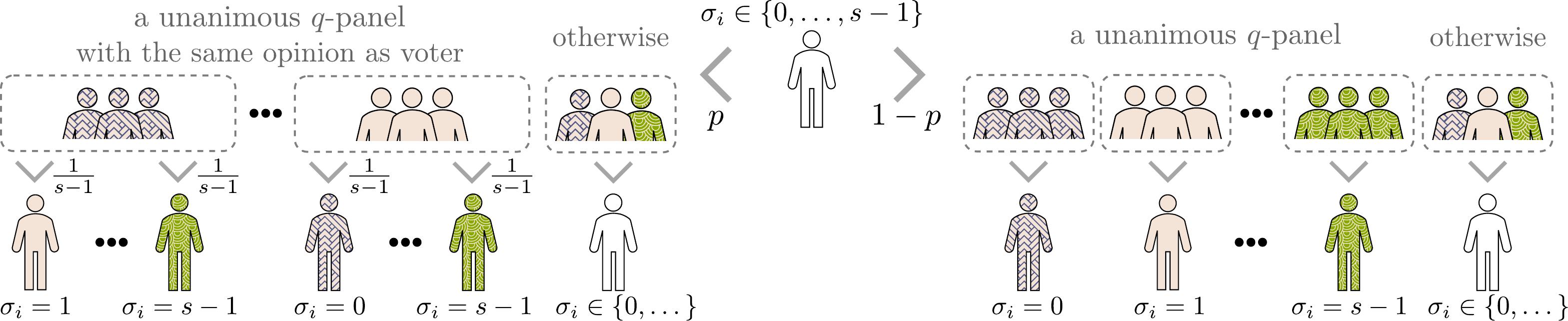}
  \caption{Visualization of the elementary update for the multi-state $q$-voter model with anticonformity. Within the annealed approach, a voter can anticonform or conform to the group of influence with complementary probabilities $p$ and $1-p$. Withing the quenched approach, a fraction $p$ of all voters are permanently anticonformists, whereas others are always conformists.}
  \label{fig:model}
\end{figure*}

The model, which we refer to as the multi-state $q$-voter model with anticonformity, is defined as follows. There is a system of $N$ voters, placed in the nodes of an arbitrary graph (here we focus on CG). Each voter $i=1,\ldots,N$ can be in one of $s$ possible states $\sigma_{i} = \alpha \in \{0,1,2,3,\dots,s-1\}$. As in the original $q$-voter model, a voter can be influenced by $q$ neighboring agents only if they are unanimous \cite{Cas:Mun:Pas:09}. As usually, we use a random sequential updating and a unit of time ($t \rightarrow t + 1$) is defined as $N$ elementary updates of duration $\Delta t$, i.e. $N\Delta t = 1$, which corresponds to one Monte Carlo step (MCS). An elementary update (schematically presented in Fig. \ref{fig:model}) consists of: (1) choosing randomly voter $i$, (2) choosing randomly a group of  $q$ neighbors of $i$, (3) checking if all $q$ neighbors are in the same state to form a group of influence, (4) updating the state of voter $i$. The last step of an update depends on considered approach, annealed or quenched.

Within the annealed approach, all agents are identical: with probability $p$ an active voter acts as an anticonformist, and with complementary probability $1-p$ as a conformist. Within the quenched approach, the system consists of two types of agents: each voter is set to be permanently anticonformist with probability $p$ or conformist with complementary probability $1-p$. If the active voter is an anticonformist and all $q$ neighbors are in the same state as the state of an active voter, it changes its state to any other, randomly chosen from the remaining equally probable $s-1$ states. On the other hand, if the active voter is a conformist and all $q$ neighbors are in the same state, the active voter copies their state. 

To describe the system on the macroscopic scale, we introduce a random variable $c_{\alpha}$ describing the concentration of agents having opinion $\alpha$:
\begin{equation}
  c_{\alpha} = \frac{N_{\alpha}}{N}, \quad \mbox{and} \quad \sum\limits_{\alpha=0}^{s-1}c_{\alpha} = 1,
  \label{eq:sumuptoone}
\end{equation}
where $N_{\alpha}$ is the number of voters in a given state. Because we use the random sequential updating, $c_{\alpha}$ can change only by $\pm 1/N$ with the respective transition rates:
\begin{align}
  \gamma^{+}(c_{\alpha}) &= Pr\left\{c_{\alpha}(t + \Delta t) = c_{\alpha}(t) + \frac{1}{N}\right\}, \nonumber \\
  \gamma^{-}(c_{\alpha}) &= Pr\left\{c_{\alpha}(t + \Delta t) = c_{\alpha}(t) - \frac{1}{N}\right\}.
\label{eq:gamma_gen}
\end{align}
The specific form of $\gamma^{+},\gamma^{-}$ can be easily calculated within MFA for the annealed as well as the quenched approach. Detailed calculations for the transition rates (\ref{eq:gamma_gen}), as well as other detailed calculations, to which we will refer later in this paper, can be found in supplementary material (SM). 
Although, $c_{\alpha}$ is a random variable, we can write the evolution equation of the corresponding expected value, under the realistic assumption that for $N \rightarrow \infty$ random variable $c_{\alpha}$ localizes to the expectation value and thus:
\begin{equation}
  \frac{dc_{\alpha}}{dt} = \gamma^{+}(c_{\alpha}) - \gamma^{-}(c_{\alpha}) = F(c_{\alpha}),
  \label{eq:dcdtgammas}
\end{equation} 
where $F(c_{\alpha})$ can be interpreted as the effective force acting on the system \cite{Nyc:Szn:Cis:12}. Because in this paper we focus on PTs, we are not interested in the temporal evolution of the system, but only in the stationary states:
\begin{equation}
  \frac{dc_{\alpha}}{dt} = 0.
  \label{eq:stationaryeq}
\end{equation} 
Due to equivalence of all opinions, one can claim that the only possible symmetry breaking schemes are those with at most two distinct stationary values \cite{Now:Sto:Szn:21, Man:Tag:20, Lin:Tay:93}. If initially several (one, two or more) opinions are equinumerous and dominate over all the others, the system reaches an absorbing state in which these opinions still dominate and are equinumerous. At the same time, all remaining opinions become equinumerous. Based on this observation, confirmed by Monte Carlo simulations, and the normalization condition (\ref{eq:sumuptoone}) we are able to write down all solutions in terms of a single arbitrarily chosen state, denoted by $c$:
\begin{align}
    c_0 &= \dots = c_{s-(\xi + 1)} = c, \nonumber \\ 
    c_{s-\xi} &= \dots = c_{s-1} = \frac{1-(s-\xi)c}{\xi},
    \label{eq:xidenoteannealed}
\end{align}
where $\xi = 1,2,3,\dots,s-1$. By $\xi = 0$ we denote the solution, where all states are equinumerous $c_0 = c_1 = \dots = c_{s-1} = 1/s$.
Knowing the above, we can determine the stationary states of the annealed and the quenched version of the model.

\begin{figure*}
  \includegraphics[width=\linewidth]{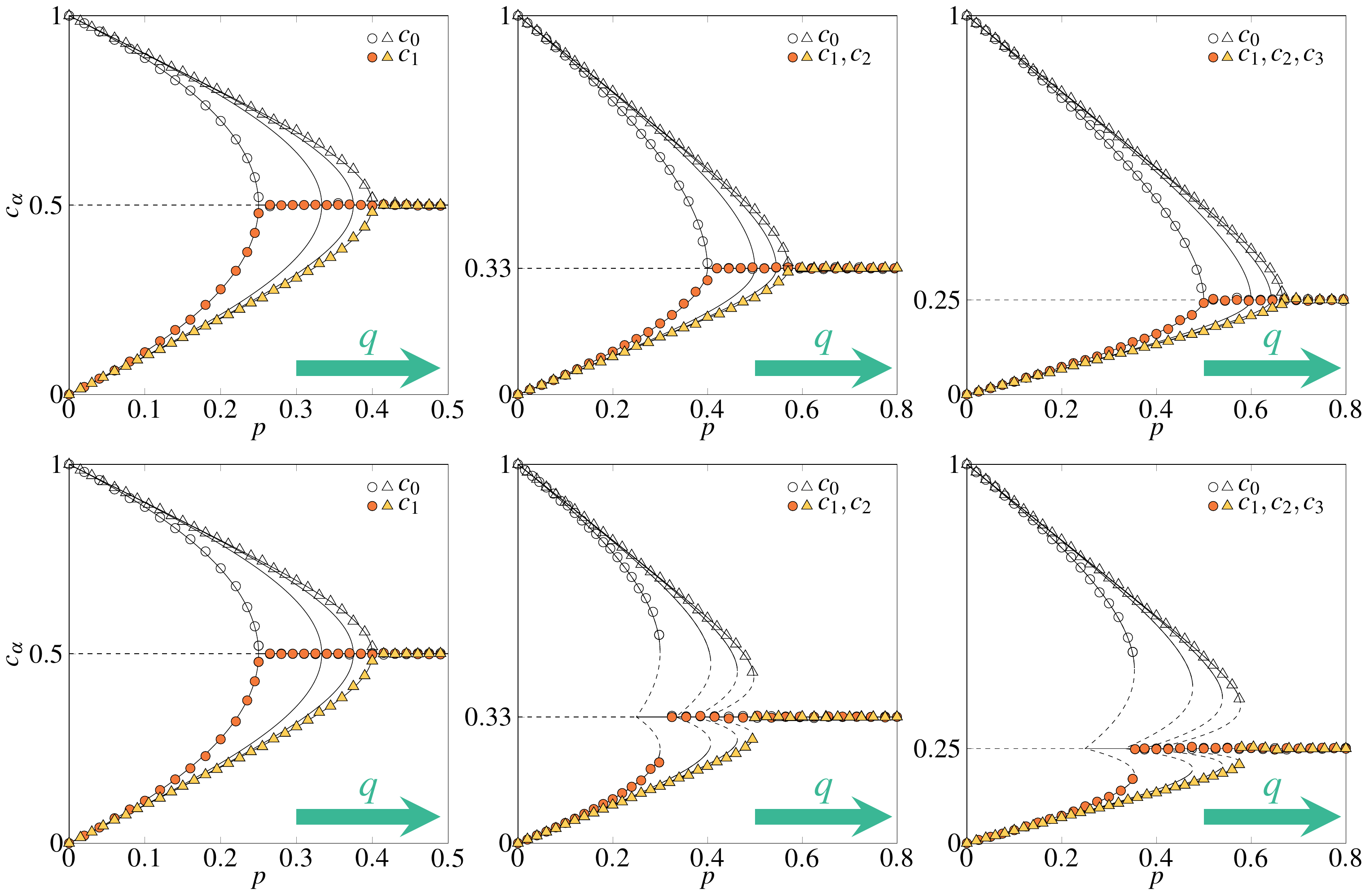}
  \caption{Stationary concentration of agents in a given state as a function of the probability of anticonformity $p$ within the annealed (upper panels) and quenched (bottom panels) approach for different values of the influence group size $q = \{2,3,4,5\}$ (changing from left to right as indicated by arrows). The number of states $s = 2$ (left column), $s=3$ (middle column) and $s=4$ (right column). Lines represent analytical results: solid and dashed lines correspond to stable and unstable steady states, respectively. Symbols represent the outcome of MC simulations for $q=\{2,5\}$ and the system size $N = 5 \times 10^5$ performed from initial condition $c_{0}=1$. The results are averaged over $10$ runs and collected after $t=5 \times 10^4$ MCS.}
  \label{fig:cp}
\end{figure*}

Under the annealed approach on the complete graph, Eq. (\ref{eq:dcdtgammas}) takes the form (see SM):
\begin{equation}
  \frac{dc_{\alpha}}{dt} = -pc_{\alpha}^{q+1} + p\sum_{i \neq \alpha} \left[ \frac{c_{i}^{q+1}}{s-1} \right] + (1-p)\sum_{i \neq \alpha} \left[ c_{i}c_{\alpha}^q - c_{\alpha} c_{i}^q \right].
  \label{eq:dcdtnoxi}
\end{equation}
Combining Eq. (\ref{eq:xidenoteannealed}) with Eqs. (\ref{eq:stationaryeq}) and (\ref{eq:dcdtnoxi}) we obtain
\begin{widetext}
  \begin{equation} 
      p = \frac{c^{q} - (s-\xi)c^{q+1} - \xi c \left(\frac{1-(s-\xi)c}{\xi}\right)^{q}}{c^q -(s-\xi)c^{q+1} - \xi c \left(\frac{1-(s-\xi)c}{\xi}\right)^{q} - \frac{\xi}{s-1}\left[\left(\frac{1-(s-\xi)c}{\xi}\right)^{q+1} - c^{q+1}\right]}.
      \label{eq:pxiannealed}
  \end{equation}
\end{widetext}
Based on the value of $\xi$ we are able to recover $s$ stationary solutions. Some of them are stable and some unstable. We can determine the stability, by looking at the sign of the derivative of the effective force $\frac{dF(c_{\alpha})}{dc}$, see SM for details. It turns out that only two types of stable stationary states of the system are possible: (\textbf{disordered}) concentrations of all opinions are identical $c_{\alpha}=1/s$ for all $\alpha$ and (\textbf{ordered}) in which the symmetry between opinions is broken and and one opinion dominates over the others. As seen in Fig.~\ref{fig:cp}, in the annealed case this order-disorder PT is continuous for all values of model's parameters $q$ and $s$.

Under the quenched approach the system  consists of two types of agents, who respond differently to group influence. Therefore, we must consider these two groups separately and write one evolution equation for the concentration $c_{(\mathbf{A},\alpha)}$ of anticonformists in state $\alpha$, and the second one for concentration $c_{(\mathbf{C},\alpha)}$ of conformists in this state \cite{Jed:Szn:17,Jed:Szn:20}. 
This will ultimately allow to obtain the total concentration of voters in state $\alpha$:
\begin{equation}
  c_{\alpha} = p c_{(\mathbf{A},\alpha)} + (1-p)c_{(\mathbf{C},\alpha)}
  \label{eq:quenchedconcentrationalpha}
\end{equation}
and the evolution of the system is given by two equations
\begin{align}
  \frac{dc_{(\mathbf{A},\alpha)}}{dt} &= - c_{(\mathbf{A},\alpha)} c_{\alpha}^q + \sum_{i \neq \alpha} \left[\frac{c_{(\mathbf{A},i)} c_{i}^q}{s-1} \right], \nonumber \\
  \frac{dc_{(\mathbf{C},\alpha)}}{dt} &= \sum_{i \neq \alpha} \left[ c_{(\mathbf{C},i)}c_{\alpha}^q - c_{(\mathbf{C},\alpha)}c_{i}^q\right].
  \label{eq:dcdtnoxiquenched}
\end{align}

By performing analogous reasoning to the annealed approach, we compute the stationary states

\begin{widetext}
  \begin{equation}
  p = \frac{\left(\frac{1-(s-\xi)c}{\xi}\right)^{q}c^{q} \left[cs^2 - (1+2c\xi)(s-\xi)\right] + \xi c^{2q} \left[c(s-\xi) - 1\right] + c\xi(s-\xi)\left(\frac{1-(s-\xi)c}{\xi}\right)^{2q}}{\xi \left(\frac{1-(s-\xi)c}{\xi}\right)^{2q} - \xi c^{2q}}
  \label{eq:pxiquenched}
\end{equation}
\end{widetext}

and determine their stability (see the SM for details). We again obtain a phase transition between the \textbf{disordered} state, in which concentrations of all opinions are identical $c_{\alpha}=1/s$ and the \textbf{ordered} state, in which one opinion dominates. However, in contrast to the annealed case, this time for $s>2$ this transition is discontinuous, as shown in the bottom panels of Fig. \ref{fig:cp}. For $s=2$, the results for the annealed and quenched approach are identical, as already shown in \cite{Jed:Szn:17}. As expected, the Monte Carlo simulations on the CG produced the same results as the analytical MFA, see Fig. \ref{fig:cp}.

\section{Results}
\begin{figure*}[!htb]
  \includegraphics[width=\linewidth]{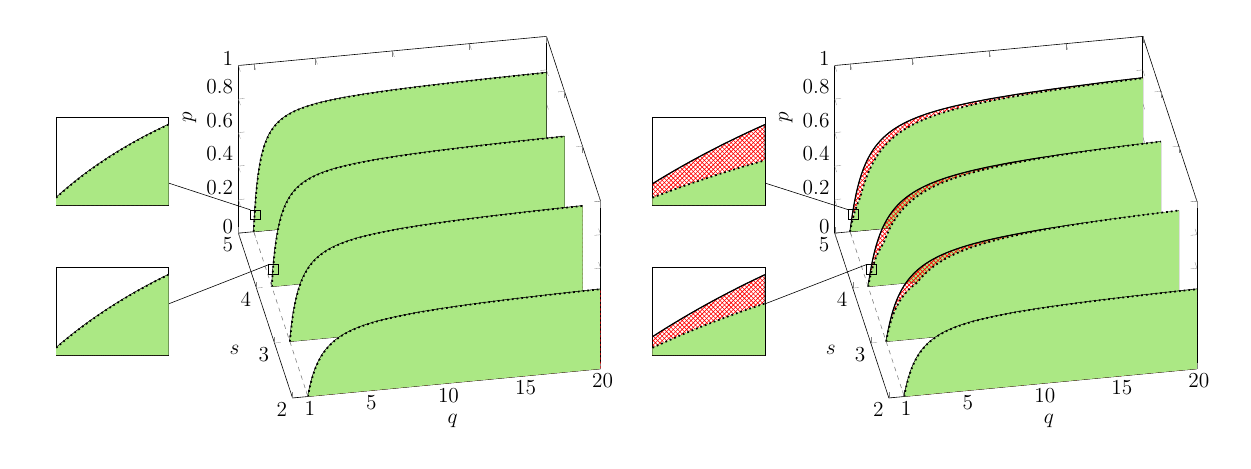}
  \caption{Phase diagrams obtained within MFA for the multi-state $q$-voter model under the annealed (left panels) and the quenched (right panels) approach. The ordered phases are marked by solid fill-color (green). The coexistence regions are marked by a crosshatched pattern (red). The disordered phases are shown as no-fill-color regions (white). Lower and upper spinodals are marked by dotted and solid lines respectively.}
  \label{fig:phasediagram}
\end{figure*}

For the binary $q$-voter model with anticonformity, that is $s=2$, the quenched approach gives the same result as the annealed one, and the phase transitions are continuous regardless of $q$, as shown in the left panels of Fig. \ref{fig:cp}. This result has already been obtained in the previous paper \cite{Jed:Szn:17} and here appears only as a special case of the general multi-state model. The new results concern $s>2$, for which unexpectedly the quenched model induces discontinuous phase transitions. While in the annealed version the phase transitions are still continuous, the quenched model displays discontinuous transitions already for $q>1$, see Figs. \ref{fig:cp} and \ref{fig:phasediagram}. For all values of the model parameters, the Monte Carlo results overlap analytical ones, as shown in Fig. \ref{fig:cp}, which was expected due to the structure of the complete graph. 

As seen in Fig. \ref{fig:phasediagram} for the fixed value of $s>2$ the size of the hysteresis, that is, the area in which ordered phase coexists with disordered one and thus the final state depends on the initial one, depends non-monotonically on the size of the influence group $q$. Initially, it increases with $q$ and reaches the maximum value at
\begin{equation}
q = \frac{s}{s-2}, 
\end{equation}
what can be calculated analytically, as shown in SM.

\section{Conclusions}
The initial inspiration for this research came from social science and was specifically related to the question of factors that influence the emergence of discontinuous phase transitions in social systems. This question with respect to models of social dynamics has been asked before in several papers \cite{Vie:Cro:16,Li:etal:16,Enc:etal:18,Iac:19,Now:Sto:Szn:21}. One might wonder why discontinuous phase transitions are relevant to social systems at all. In fact, they are important because it turns out that hysteresis and critical mass, which are indicators of discontinuous phase transitions, are empirically observed in real social systems \cite{Str:Liz:17,Cen:etal:18,Gui:Bar:Cen:21}. The importance of discontinuous phase transitions was one of the reasons why, for example, the $q$-voter with independence was studied more intensively than the $q$-voter model with anticonformity \cite{Jed:17,Vie:etal:20,Gra:Kra:20,Chm:etal:20,Civ:21,Jan:Chm:22}.

Within the annealed approach, the $q$-voter model with anticonformity displays only continuous PT, independently of the number of states $s$ and the size of the influence source $q$. On the contrary, the $q$-voter model with independence shows discontinuous PTs under the annealed approach above the tricritical point $q^*(s)$, where $q^*(2)=5$ \cite{Nyc:Szn:Cis:12,Jed:Szn:17} and $q^*(s>2)=1$ \cite{Now:Sto:Szn:21}. Moreover, it was shown that for the $q$-voter with independence replacing the annealed disorder by the quenched one kills discontinuous phase transitions for $s=2$ \cite{Jed:Szn:17} or rounds them for $s>2$ \cite{Now:Sto:Szn:21}. These previous results were in agreement with the state of the art \cite{Imr:Ma:75,Aiz:Weh:89,Voj:06,Mun:etal:10,Bor:Mar:Mun:13,Vil:Bon:Mun:14}. On the contrary, in this paper we have shown that the opposite phenomenon can also be observed. 

We are aware that obtaining the same results independently within the two methods (analytical and Monte Carlo simulations) does not mean that we understand the observed phenomenon. Unfortunately, heuristic understanding is still lacking. Nevertheless, we have decided to present these results, hoping for the help of the readers. Admittedly, nowhere in the literature have we found a proof that the phenomenon we observe is impossible for a complete graph. On the other hand, we have not found any paper in which anyone has observed such a phenomenon. From this perspective, the model studied here should be treated as an example that shows that a quenched disorder can support discontinuous phase transitions in some cases. Therefore, we believe that our finding goes far beyond social physics and should be interesting to a broad audience.

\section*{Author contributions}
B.N. was responsible for all analytical calculations and Monte Carlo simulations. K. S-W. was responsible for supervising the research and funding acquisition. Both authors were writing, reviewing, and editing the manuscript.

\section*{Acknowledgments}
This work was partially supported by funds from the National Science Center (NCN,Poland) through Grant 2019/35/B/HS6/02530.

\end{document}


\preprint{APS/123-QED}

\title{Supplementary material for\\Switching from a continuous to discontinuous phase transition under the quenched disorder.}

\author{Bartłomiej Nowak}
  \email{bartlomiej.nowak@pwr.edu.pl}
\author{Katarzyna Sznajd-Weron}
  \email{katarzyna.weron@pwr.edu.pl}
\affiliation{
Department of Theoretical Physics, Wrocław University of Science and Technology, 50-370 Wrocław, Poland
}
\date{\today}

\begin{abstract}
In this supplementary material, we will provide the details of the mathematical calculations for the $q$-voter model with anticonformity under two types of disorder: quenched and annealed. 
\end{abstract}

\maketitle

\section{Derivation of the model}

\begin{figure*}[!htb]
  \includegraphics[width=\linewidth]{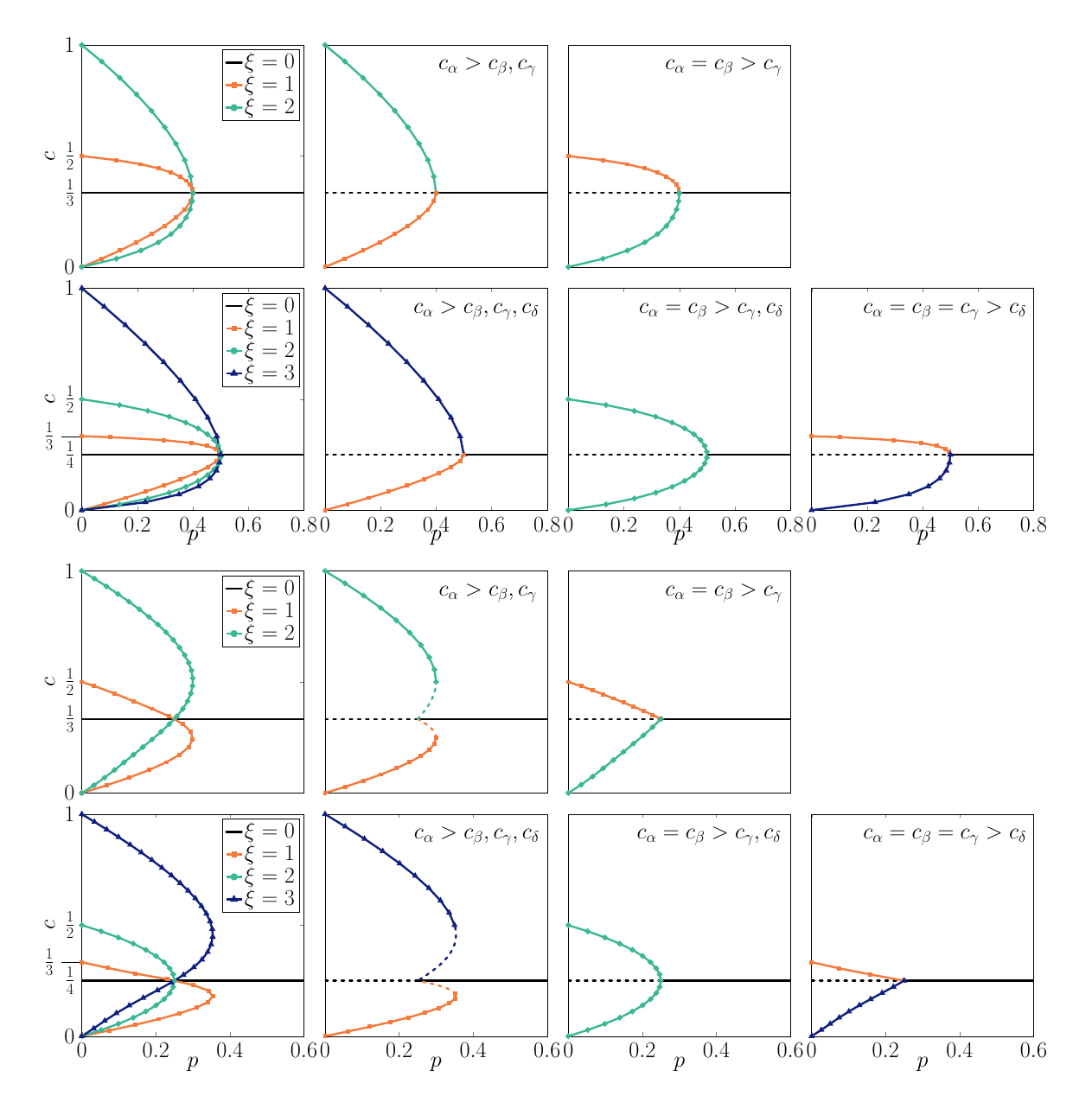}
  \caption{Steady states for $q = 2$ within the annealed (two upper rows) as well as the quenched approach (two bottom rows) for: $s = 3$ (the shorter rows, i.e. the first and third), $s = 4$ (the longer rows, i.e. the second and fourth). Plots in the first column show all possible solutions indexed  $\xi$ without the distinction between the stable and unstable ones obtained from Eq. (\ref{eq:pxiannealed}) for the annealed approach and from Eq. (\ref{eq:pxiquenched}) for the quenched one. The remaining columns represent stationary states for initial conditions indicated in the top right corners of each panel, where $\alpha, \beta, \gamma,\delta \in \{0,1,\dots,s-1\}$. Stable solutions are marked with the solid lines with symbols, and unstable with the dashed lines.}
  \label{fig:ksisplit}
\end{figure*} 

\begin{figure*}[!htb]
  \includegraphics[width=\linewidth]{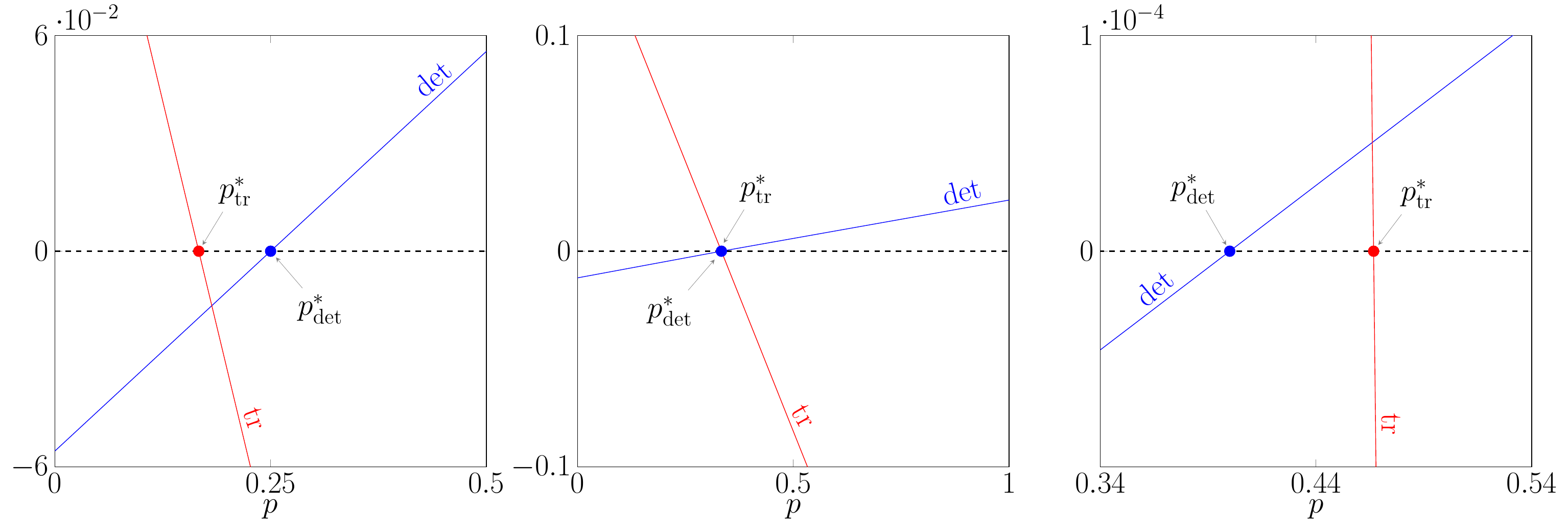}
  \caption{Dependence between the determinant (\ref{eq:determinant}), trace (\ref{eq:trace}) and value of the parameter $p$ for different number of states and size of the influence group: $s = 3$ and $q = 2$ (left); $s = 3$ and $q = 3$ (middle); $s = 3$ and $q = 5$ (right). Lines indicate values of the trace ($\operatorname{tr}$) and determinant ($\det$) denoted above each line. Points describe the values of $p$ at which trace and determinant change sign, see Eqs. (\ref{eq:trroot}) and (\ref{eq:detroot}).}
  \label{fig:tracedetdiff}
\end{figure*}

\subsection{\label{app:Derivationannealed} Annealed approach}
The total concentration of voters in a state $\alpha$ can increase by $1/N$ only if we pick an active voter in a different state than $\alpha$. Then the active voter can anticonform with probability $p$ or conform with probability $1-p$. In both cases, a lobby of $q$ neighbors is chosen randomly (without repetitions) and it can change the state of an active voter only if it is unanimous. 

To present the detailed calculation, we introduce the following notation:
\begin{itemize}
    \item $P(i)$ -- the probability of choosing randomly an active (i.e., the one whose state will be updated) voter in state $i = \{0,1,\dots,s-1\}$. 
    \item $P(\alpha|i)$ -- the conditional probability of picking a neighbor in state $\alpha$ given that an active voter is in state $i$.
\end{itemize}

Within the above notation, the probability of choosing $q$ neighbors in state $\alpha$ is equal to $P^{q}(\alpha|i)$. In the case of conformity, $\alpha$ has to be different from $i$ to change the state of an active voter. On the other hand, in the case of anticonformity, $\alpha=i$ is needed to change the state of an active voter, and then state $\alpha$ must be chosen. The latter occurs with probability $1/(s-1)$, because a voter chooses a new state randomly from $s-1$ states.

Analogous reasoning can be carried out for the situation in which the total concentration of voters in a state $\alpha$ decreases. In result, the transition rates can be expressed explicitly as:
\begin{align}
  \gamma^{+}(c_{\alpha}) &= \sum_{i \neq \alpha} P(i) \left[p \frac{P^{q}(i|i)}{s-1} + (1-p) P^{q}(\alpha|i) \right] \nonumber \\ 
  \gamma^{-}(c_{\alpha}) &= \sum_{i \neq \alpha} P(\alpha) \left[p \frac{P^{q}(\alpha|\alpha)}{s-1} + (1-p)P^{q}(i|\alpha) \right] 
\label{eq:gamma}
\end{align}

Within MFA, the events of picking an active voter in a given state $i$ and a neighbor in a state $\alpha$ are independent, so all conditional probabilities $P(\alpha|i)$, $P(i|i)$ are equal to $P(\alpha)$, $P(i)$ respectively. Moreover, the local concentration of agents is the state $\alpha$ or $i$ is equal to the global one, which means that $P(\alpha) = c_{\alpha}$ and $P(i) = c_{i}$. Therefore, within MFA Eq. (\ref{eq:gamma}) boils down to:
\begin{align}
  \gamma^{+}(c_{\alpha}) &= \sum_{i \neq \alpha} c_{i} \left[p \frac{c_{i}^q}{s-1} + (1-p) c_{\alpha}^q \right] \nonumber \\ 
  \gamma^{-}(c_{\alpha}) &= \sum_{i \neq \alpha} c_{\alpha} \left[p \frac{c_{\alpha}^q}{s-1} + (1-p)c_{i}^q \right].
\label{eq:gamma_mfa}
\end{align}

This leads to:
\begin{align}
  \frac{dc_{\alpha}}{dt} &= \gamma^+(c_{\alpha}) - \gamma^-(c_{\alpha}) \nonumber \\
  & = -pc_{\alpha}^{q+1} + p\sum_{i \neq \alpha} \left[ \frac{c_{i}^{q+1}}{s-1} \right] + (1-p)\sum_{i \neq \alpha} \left[ c_{i}c_{\alpha}^q - c_{\alpha} c_{i}^q \right].
  \label{eq:dcdtnoxi}
\end{align}
Stationary states can be found by solving the following equation
\begin{equation}
    \frac{dc_{\alpha}}{dt} = 0.
    \label{eq:stationary}
\end{equation}
Recalling the notation
\begin{align}
    c_0 &= \dots = c_{s-(\xi + 1)} = c, \nonumber \\ 
    c_{s-\xi} &= \dots = c_{s-1} = \frac{1-(s-\xi)c}{\xi},
    \label{eq:annealedksi}
\end{align}
Eq. (\ref{eq:stationary}) takes form
\begin{align}
    &\frac{p\xi}{s-1}\left[\left(\frac{1-(s-\xi)c}{\xi}\right)^{q+1} - c^{q+1}\right] + (1-p)c^{q} \nonumber \\ 
    &- (1-p)\left[(s-\xi)c^{q+1} + \xi c \left(\frac{1-(s-\xi)c}{\xi}\right)^q\right] = 0.
\end{align}
Solving it for $p$ yields the stationary solution 
\begin{widetext}
  \begin{equation} 
      p = \frac{c^{q} - (s-\xi)c^{q+1} - \xi c \left(\frac{1-(s-\xi)c}{\xi}\right)^{q}}{c^q -(s-\xi)c^{q+1} - \xi c \left(\frac{1-(s-\xi)c}{\xi}\right)^{q} - \frac{\xi}{s-1}\left[\left(\frac{1-(s-\xi)c}{\xi}\right)^{q+1} - c^{q+1}\right]}.
      \label{eq:pxiannealed}
  \end{equation}
\end{widetext}

From equation (\ref{eq:pxiannealed}) one can reproduce all $s$ stationary solutions, see two upper rows in Fig. \ref{fig:ksisplit}.

\subsection{\label{app:Derivationquenched} Quenched approach}

Within the quenched approach, two groups of agents exist: conformists and anticonformists. Therefore, the separate reasoning has to be provided within each group and the dynamics is given by two equations for concentration of agents in state $\alpha$ among anticonformists $c_{(\mathbf{A},\alpha)}$ and conformists $c_{(\mathbf{C},\alpha)}$
\begin{align}
    \frac{dc_{(\mathbf{A},\alpha)}}{dt} &= \gamma^{+}_{\mathbf{A}}(c_{(\mathbf{A},\alpha)}) - \gamma^{-}_{\mathbf{A}}(c_{(\mathbf{A},\alpha)}) = F_{\mathbf{A}}(c_{(\mathbf{A},\alpha)}), \\
    \frac{dc_{(\mathbf{C},\alpha)}}{dt} &= \gamma^{+}_{\mathbf{C}}(c_{(\mathbf{C},\alpha)}) - \gamma^{-}_{\mathbf{C}}(c_{(\mathbf{C},\alpha)}) = F_{\mathbf{C}}(c_{(\mathbf{C},\alpha)}),
\label{eq:dcdtquenched}
\end{align}
where $\gamma^{+}_{\mathbf{A}}(c_{(\mathbf{A},\alpha)})$, $\gamma^{-}_{\mathbf{A}}(c_{(\mathbf{A},\alpha)})$ denotes transition rates for anticonformist and $\gamma^{+}_{\mathbf{C}}(c_{(\mathbf{C},\alpha)})$, $\gamma^{-}_{\mathbf{C}}(c_{(\mathbf{C},\alpha)})$ for conformist agents. Forces acting on the system for anticonformist and conformist agents are denoted with $F_{\mathbf{A}}(c_{(\mathbf{A},\alpha)})$ and $F_{\mathbf{C}}(c_{(\mathbf{C},\alpha)})$, respectively.

The concentration of anticonformists in state $\alpha$ can increase by $1/N$ only if an anticonformist in state $i \neq \alpha$ is randomly chosen to be an active voter. On the other hand, the concentration of conformists in state $\alpha$ can increase by $1/N$ only if a conformist in state $i \neq \alpha$ is randomly chosen to be an active voter. Therefore, to derive the explicit form of transition rates, 
we keep notation from \ref{app:Derivationannealed}, and additionally we introduce:
\begin{itemize}
    \item $P_{\mathbf{A}}(i)$ -- the probability of choosing  anticonformist in state $i = \{0,1,\dots,s-1\}$ to be an active voter.
    \item $P_{\mathbf{C}}(i)$ -- the probability of choosing  conformist in state $i = \{0,1,\dots,s-1\}$ to be an active voter.
\end{itemize}

\begin{figure}[!hb]
  \includegraphics[width=\linewidth]{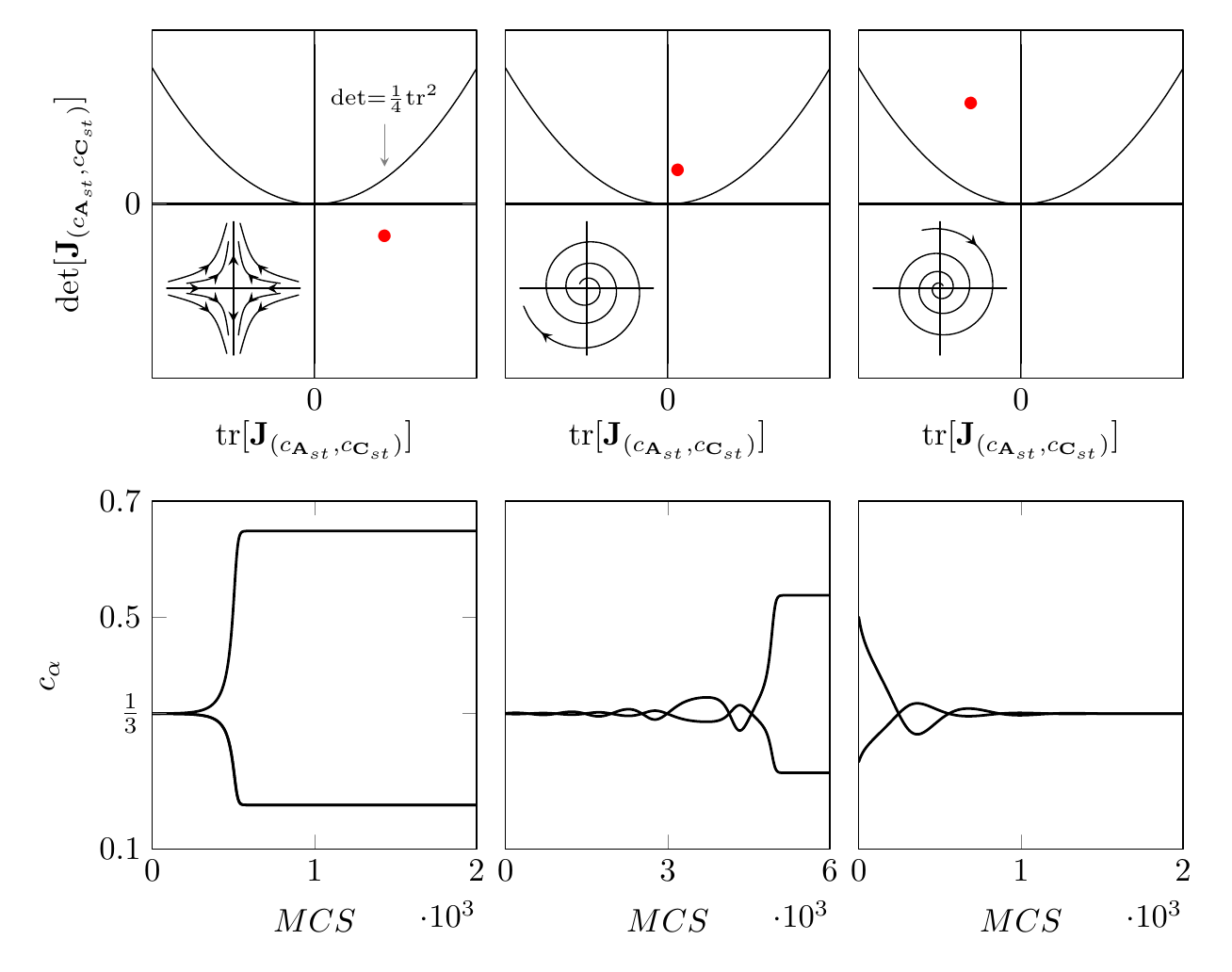}
  \caption{The dependence between determinant and trace (Poincar\' e diagrams, upper panels) and sample trajectories (bottom panels) for parameters $q = 5$ and $s = 3$, and three values of $p$: $p = 0.35$ (left column), $p = 0.45$ (middle column), $p = 0.55$ (right column). Dots in Poincar\' e diagrams correspond to values of determinant and trace for the solution $(c_{\mathbf{A}},c_{\mathbf{C}}) = (1/s, 1/s)$. Stability of this solution depends on $p$, which is shown by flow diagrams in the left corners of upper panels: negative determinant implies unstable saddle solution (left column), positive determinant and positive trace above critical parabola implies unstable spiral (middle column), positive determinant and negative trace above critical parabola implies stable spiral (right column).}
  \label{fig:trajectories}
\end{figure}

Within such a notation, the transition rates can be expressed explicitly as
\begin{align}
  \gamma^{+}_{\mathbf{A}}(c_{(\mathbf{A},\alpha)}) &= \sum_{i \neq \alpha}\left[ \frac{P_{\mathbf{A}}(i) P^{q}(i|i)}{s-1}\right], \nonumber \\
  \gamma^{-}_{\mathbf{A}}(c_{(\mathbf{A},\alpha)}) &= \sum_{i \neq \alpha}\left[ \frac{P_{\mathbf{A}}(\alpha) P^{q}(\alpha|\alpha)}{s-1}\right], \nonumber \\
  \gamma^{+}_{\mathbf{C}}(c_{(\mathbf{C},\alpha)}) &= \sum_{i \neq \alpha}\left[ P_{\mathbf{C}}(i) P^{q}(\alpha|i)\right], \nonumber \\
  \gamma^{-}_{\mathbf{C}}(c_{(\mathbf{C},\alpha)}) &= \sum_{i \neq \alpha}\left[ P_{\mathbf{C}}(\alpha) P^{q}(i|\alpha)\right].
\end{align}

Within MFA $P(\alpha|\beta)=P(\alpha)$, $P(\alpha) = c_{\alpha}$, $P_{\mathbf{A}}(\alpha) = c_{(\mathbf{A},\alpha)}$ and $P_{\mathbf{C}}(\alpha) = c_{(\mathbf{C},\alpha)}$ and thus:
\begin{align}
  \gamma^{+}_{\mathbf{A}}(c_{(\mathbf{A},\alpha)}) &= \sum_{i \neq \alpha}\left[ \frac{c_{(\mathbf{A},i)} c_{i}^q}{s-1}\right], \nonumber \\
  \gamma^{-}_{\mathbf{A}}(c_{(\mathbf{A},\alpha)}) &= \sum_{i \neq \alpha}\left[ \frac{c_{(\mathbf{A},\alpha)} c_{\alpha}^q}{s-1}\right], \nonumber \\
  \gamma^{+}_{\mathbf{C}}(c_{(\mathbf{C},\alpha)}) &= \sum_{i \neq \alpha}\left[ c_{(\mathbf{C},i)}c_{\alpha}^q\right], \nonumber \\
  \gamma^{-}_{\mathbf{C}}(c_{(\mathbf{C},\alpha)}) &= \sum_{i \neq \alpha}\left[ c_{(\mathbf{C},\alpha)}c_{i}^q\right].
\end{align}

and in result the model can be described as follows
\begin{align}
  \frac{dc_{(\mathbf{A},\alpha)}}{dt} &= - c_{(\mathbf{A},\alpha)} c_{\alpha}^q + \sum_{i \neq \alpha} \left[\frac{c_{(\mathbf{A},i)} c_{i}^q}{s-1} \right],\nonumber \\
  \frac{dc_{(\mathbf{C},\alpha)}}{dt} &= \sum_{i \neq \alpha} \left[ c_{(\mathbf{C},i)}c_{\alpha}^q - c_{(\mathbf{C},\alpha)}c_{i}^q\right]
  \label{eq:dcdtnoxiquenched}
\end{align}
and the total concentration of agents in the state $\alpha$ is given by
\begin{equation}
c_{\alpha} = p c_{(\mathbf{A},\alpha)} + (1-p)c_{(\mathbf{C},\alpha)}.
\label{eq:totalconcquench}
\end{equation}

Similarly to annealed version, we can express all stationary concentrations by a concentration of arbitrarily chosen state denoted with $c$ using Eq. (\ref{eq:annealedksi}), whereas by $c_{\mathbf{A}}$, $c_{\mathbf{C}}$ we express concentrations of anticonformists and conformists in this state respectively
\begin{align}
    c_{(\mathbf{A},0)} &= \dots = c_{(\mathbf{A},s-(\xi + 1))} = c_{\mathbf{A}}, \nonumber \\
    c_{(\mathbf{A},s-\xi)} &= \dots = c_{(\mathbf{A},s-1)} = \frac{1-(s-\xi)c_{\mathbf{A}}}{\xi}; \label{eq:quenchedstationaryxi2} \\[1em]
    c_{(\mathbf{C},0)} &= \dots = c_{(\mathbf{C},s-(\xi + 1))} = c_{\mathbf{C}}, \nonumber \\
    c_{(\mathbf{C},s-\xi)} &= \dots = c_{(\mathbf{C},s-1)} = \frac{1-(s-\xi)c_{\mathbf{C}}}{\xi}.
    \label{eq:quenchedstationaryxi3}
\end{align}

\begin{figure}[!htb]
    \includegraphics[width=\linewidth]{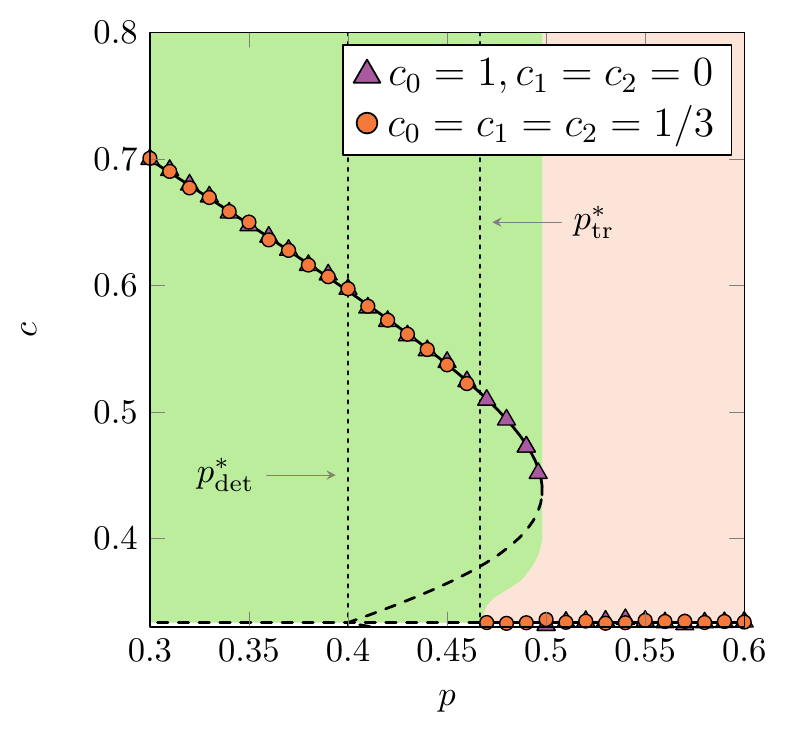}
  \caption{The dependence between the stationary concentration $c$ of agents in state $0$ and probability of anticonformity $p$ within the quenched approach for $q=5$ and $s=3$. Lines are given by Eq. (\ref{eq:pxiquenched}): solid and dashed lines correspond to stable and unstable steady states, respectively. Vertical dotted lines represent critical points $p_{\det}^{*}$ and $p_{\operatorname{tr}}^{*}$. Two color shaded areas correspond to two different basins of attraction obtained numerically from Eqs. (\ref{eq:dcdtnoxiquenched}) and (\ref{eq:totalconcquench}): (left area, green color online) trajectories converge to stationary ordered state, (right area, pink color online) trajectories converge to the stationary disordered state. Symbols represent the outcome of MC simulations for the system size $N = 5 \times 10^5$ performed from two initial conditions indicated in the legend. The results are averaged over ten runs and collected after $t=5 \times 10^4$ MCS.}
  \label{fig:hysteresis}
\end{figure}

\begin{align}
  &\frac{\xi}{s-1} \left[
    \frac{1-(s-\xi)c_{\mathbf{A}}}{\xi}\left(\frac{1-(s-\xi)c}{\xi}\right)^{q} - c_{\mathbf{A}}c^{q}
  \right] = 0 \nonumber \\
  &\xi \left[ \frac{1-(s-\xi)c_{\mathbf{C}}}{\xi}c^{q} - c_{\mathbf{C}} \left(\frac{1-(s-\xi)c}{\xi}\right)^q \right] = 0
  \label{eq:quenchedstationaryeqs}
\end{align}
and total concentration is given by
\begin{equation}
    c = p c_{\mathbf{A}} + (1-p)c_{\mathbf{C}}.
    \label{eq:totalconcquenchstationary}
\end{equation}
Eqs. (\ref{eq:quenchedstationaryeqs}) give following stationary solutions
\begin{align}
  c_{\mathbf{A}} &= \frac{\left(\frac{1-(s-\xi)c}{\xi}\right)^q}{(s-\xi)\left(\frac{1-(s-\xi)c}{\xi}\right)^q + \xi c^{q}} \nonumber \\
  c_{\mathbf{C}} &= \frac{c^q}{(s-\xi)c^{q} + \xi \left(\frac{1-(s-\xi)c}{\xi}\right)^q}
\end{align}
and combining them with (\ref{eq:totalconcquenchstationary}) we obtain
\begin{widetext}
  \begin{equation}
  p = \frac{\left(\frac{1-(s-\xi)c}{\xi}\right)^{q}c^{q} \left[cs^2 - (1+2c\xi)(s-\xi)\right] + \xi c^{2q} \left[c(s-\xi) - 1\right] + c\xi(s-\xi)\left(\frac{1-(s-\xi)c}{\xi}\right)^{2q}}{\xi \left(\frac{1-(s-\xi)c}{\xi}\right)^{2q} - \xi c^{2q}}
  \label{eq:pxiquenched}
\end{equation}
\end{widetext}
With Eq. (\ref{eq:pxiquenched}) one can reproduce all $s$ stationary solutions for the quenched q-voter model with anticonformity, see two bottom rows in Fig. \ref{fig:ksisplit}.

\section{Stability analysis}
\subsection{\label{app:SAannealed} Annealed approach}

Information about the stability of the system is given by the sign of the derivative
\begin{align}
  \frac{dF(c)}{dc} &= (1-p)qc^{q-1} - (1-p)(s-\xi)(q+1)c^{q} \nonumber \\
  &+ (1-p)cq (s-\xi)\left(\frac{1-(s-\xi)c}{\xi}\right)^{q-1} \nonumber \\
  &- (1-p)\xi\left(\frac{1-(s-\xi)c}{\xi}\right)^q - p\frac{(q+1)\xi}{s-1}c^q \nonumber \\
  &- p\frac{(q+1)}{s-1}(s-\xi)\left(\frac{1-(s-\xi)c}{\xi}\right)^q.
\label{eq:dFdc}
\end{align}
Stability analysis allows us to calculate the lower spinodal, that is, the point at which the disordered phase $c = 1/s$ loses stability. Thus, to calculate it one should check the sign of derivative (\ref{eq:dFdc}) at $c = 1/s$: 
\begin{equation}
  \left.\frac{dF(c)}{dc}\right|_{c = 1/s} = \left(\frac{1}{s}\right)^q \left[(1-p)s(q-1) - ps\frac{q+1}{s-1} \right].
\label{eq:dFdc_1s}
\end{equation}

For the fixed values of $q$ and $s$, the derivative (\ref{eq:dFdc_1s}) is equal to 0 at 
\begin{equation}
  p_{1}^{*} = \frac{(s-1)(q-1)}{(s-1)(q-1)+q+1}.
\end{equation}
The derivative given by Eq. (\ref{eq:dFdc_1s}) has a positive sign for $p < p_{1}^{*}$ (unstable solution) and a negative sign for $p > p_{1}^{*}$ (stable solution). Thus, $p_{1}^{*}$ is the lower spinodal. Since there is only a continuous PT in the annealed approach, so $p_{1}^{*}$ is simultaneously the upper spinodal.

\subsection{\label{app:SAquenched} Quenched approach}
The stability of the system under the quenched approach is determined by the signs of the determinant and the trace of the following Jacobian matrix:
\begin{equation}
\mathbf J(c_{\mathbf{A}},c_{\mathbf{C}}) = \begin{bmatrix}
  \dfrac{\partial F_{\mathbf{A}}}{\partial c_{\mathbf{A}}} & \dfrac{\partial F_{\mathbf{A}}}{\partial c_{\mathbf{C}}} \\[1em]
  \dfrac{\partial F_{\mathbf{C}}}{\partial c_{\mathbf{A}}} & \dfrac{\partial F_{\mathbf{C}}}{\partial c_{\mathbf{C}}} \end{bmatrix}.
\end{equation}
The appropriate derivatives are as follows
\begin{align}
  \frac{\partial F_{\mathbf{A}}}{\partial c_{\mathbf{A}}} ={}& \frac{qp(s-\xi)}{\xi(s-1)}\left(\frac{1-(s-\xi)c}{\xi}\right)^{q-1}\left((s-\xi)c_{\mathbf{A}} - 1\right) \nonumber \\
  & - \frac{qp\xi}{s-1}c_{\mathbf{A}}c^{q-1} 
  - \frac{s-\xi}{s-1}\left(\frac{1-(s-\xi)c}{\xi}\right)^q \nonumber \\
  & - \frac{\xi}{s-1}c^q, \\
  \frac{\partial F_{\mathbf{A}}}{\partial c_{\mathbf{C}}} ={}& \frac{q(1-p)(s-\xi)}{\xi(s-1)}\left(\frac{1-(s-\xi)c}{\xi}\right)^{q-1}((s-\xi) c_{\mathbf{A}} -1) \nonumber \\
  & - \frac{q(1-p)\xi}{s-1}c_{\mathbf{A}}c^{q-1}, \\
  \frac{\partial F_{\mathbf{C}}}{\partial c_{\mathbf{A}}} ={}& 
  qpc^{q-1}(1-(s-\xi)c_{\mathbf{C}}) \nonumber \\
  &+ qp(s-\xi)c_{\mathbf{C}}\left(\frac{1-(s-\xi)c}{\xi}\right)^{q-1}, \\
  \frac{\partial F_{\mathbf{C}}}{\partial c_{\mathbf{C}}} ={}& 
  q(1-p)c^{q-1}\left(1 - (s-\xi)c_{\mathbf{C}}\right) \nonumber \\ 
  &+ q(1-p)(s-\xi)c_{\mathbf{C}}\left(\frac{1-(s-\xi)c}{\xi}\right)^{q-1} \nonumber \\
  &- (s-\xi)c^q - \xi \left(\frac{1-(s-\xi)c}{\xi}\right)^{q}.
\end{align}

Similarly to the annealed approach, we are able to calculate the lower spinodal point. It will be given by the sign of the trace and the determinant of the stationary solution $(c_{\mathbf{A}} = 1/s, c_{\mathbf{C}} = 1/s)$
\begin{align}
    \det\left[\mathbf{J}{\left(\frac{1}{s},\frac{1}{s}\right)}\right] &= \frac{\partial F_{\mathbf{A}}}{\partial c_{\mathbf{A}}} \frac{\partial F_{\mathbf{C}}}{\partial c_{\mathbf{C}}} - \frac{\partial F_{\mathbf{A}}}{\partial c_{\mathbf{C}}} \frac{\partial F_{\mathbf{C}}}{\partial c_{\mathbf{A}}} \nonumber \\
    &= \left(\frac{1}{s}\right)^{2q} \frac{s^2}{s-1}\left[q(2p-1) + 1\right], \label{eq:determinant} \\
    \operatorname{tr}\left[\mathbf{J}{\left(\frac{1}{s},\frac{1}{s}\right)}\right] &= \frac{\partial F_{\mathbf{A}}}{\partial c_{\mathbf{A}}} + \frac{\partial F_{\mathbf{C}}}{\partial c_{\mathbf{C}}} \nonumber \\
    &= \left(\frac{1}{s}\right)^{q-1} \left[q - \frac{s(qp+1)}{s-1}\right]. \label{eq:trace}
\end{align}

From Eq. (\ref{eq:determinant} we can observe that the determinant is strictly increasing with $p$ and changes the sign from negative to positive at the point 
\begin{equation}
    p_{\det}^{*} = \frac{q-1}{2q}.
    \label{eq:detroot}
\end{equation}
On the other hand, the trace strictly decreases with $p$ and changes sign from positive to negative at the point. 
\begin{equation}
    p_{\operatorname{tr}}^{*} = 1 - \frac{1}{s} - \frac{1}{q}.
    \label{eq:trroot}
\end{equation}


The steady state is stable when the trace is negative while the determinant is positive, thus the solution $(c_{\mathbf{A}} = 1/s, c_{\mathbf{C}} = 1/s)$ is stable when $p > p_{\det}^{*}$ and $p > p_{\operatorname{tr}}^{*}$. This relation can be divided into three cases.
\begin{enumerate}
    \item Root for the determinant is bigger than root for the trace $p_{\det}^{*} > p_{\operatorname{tr}}^{*}$, see left panel in Fig. \ref{fig:tracedetdiff}. Thus lower spinodal is given by $p_{1}^{*} = p_{\det}^{*}$, because trace is already negative when determinant changes its sign to positive, i.e. the disordered state becomes stable. This case occurs for example if $s = 2$, because then regardless of $q$ the condition $p_{\det}^{*} > p_{\operatorname{tr}}^{*}$ is fulfilled.
    
    \item Roots $p_{\det}^{*}$ and $p_{\operatorname{tr}}^{*}$ are equal and lower spinodal is given by $p_{1}^{*} = p_{\det}^{*} = p_{\operatorname{tr}}^{*}$, see middle panel in Fig. \ref{fig:tracedetdiff}. This happens when the following relation is fulfilled
        \begin{equation}
              q = \frac{s}{s-2}. 
              \label{eq:tracedetequal}
        \end{equation}
        The only possible integer solutions are $(s = 3, q = 3)$ and $(s = 4, q = 2)$.
    \item Root for the trace is bigger than root for the determinant $p_{\operatorname{tr}}^{*} > p_{\det}^{*}$, see right panel in Fig. \ref{fig:tracedetdiff}. In this case, we can point out three regions:
    \begin{enumerate}
        \item For $p < p_{\det}^{*}$  the determinant is negative and in the result the disordered solution is the saddle point, see left panels in Fig. \ref{fig:trajectories}. 
        \item 
        For $p_{\det}^{*} < p < p_{\operatorname{tr}}^{*}$  both determinant and trace are positive, thus the disordered solution is a source and in most cases it is a spiral source (unstable spiral). This spiral source correspond to spreading oscillations which reach attractive stable solution different from $1/s$, see middle panels in Fig. \ref{fig:trajectories}.
        \item For $p > p_{\operatorname{tr}}^{*}$ determinant is positive while trace is negative, so the disordered state is a sink and in most cases it is a spiral sink (stable spiral). This spiral sink correspond to damped oscillations which reach stable solution equals to $1/s$, see right panels in Fig. \ref{fig:trajectories}. Hence, the disordered solution become stable when $p > p_{\operatorname{tr}}^{*}$, see Fig. \ref{fig:hysteresis}.
        \end{enumerate}

\end{enumerate}